\documentclass[aps,prl,showpacs,amsmath,amssymb,amsfonts,lengthcheck,twocolumn,longbibliography,superscriptaddress]{revtex4-2}
\usepackage{epsfig,graphicx,graphics,amsmath,amssymb,float}
\usepackage[T1]{fontenc}
\usepackage[latin9]{inputenc}
\usepackage{amsmath}
\usepackage{amssymb}
\usepackage{xcolor}
\usepackage{amscd}
\usepackage{bm}
\usepackage{bbold}
\usepackage{psfrag}
\usepackage{mathrsfs}
\usepackage{float}

\usepackage{ulem}

\usepackage{bbm} 

\usepackage{chemformula}
\usepackage[caption=false]{subfig}
\usepackage{subfig}
\usepackage{color}
\usepackage[bookmarks=true,colorlinks,linkcolor=blue,urlcolor=blue,citecolor=blue]{hyperref}

\usepackage{dutchcal}
\DeclareMathAlphabet{\matholdcal}{OMS}{cmsy}{m}{n}

\newcommand{\be}{\begin{equation}}
\newcommand{\ee}{\end{equation}}
\newcommand{\bea}{\begin{eqnarray}}
\newcommand{\eea}{\end{eqnarray}}

\newcommand{\mc}{\matholdcal}
\newcommand{\ma}{\mathrm}

\newcommand{\te}{\text}

\begin{document}


\title{Decay of excess work in one-dimensional quantum critical systems}

\author{Moallison F. Cavalcante}
\email{moall@unicamp.br}
\affiliation{
Gleb Wataghin Physics Institute, The University of Campinas, 13083-859, Campinas, S\~{a}o Paulo, Brazil}
\author{Fl\'{a}via B. Ramos}
\affiliation{
International Institute of Physics, Universidade Federal do Rio Grande do Norte, Natal, RN, 59078-970, Brazil}
\author{Rodrigo G. Pereira}
\affiliation{
International Institute of Physics, Universidade Federal do Rio Grande do Norte, Natal, RN, 59078-970, Brazil}
\author{ Marcus V. S. Bonan\c{c}a}
\affiliation{
Gleb Wataghin Physics Institute, The University of Campinas, 13083-859, Campinas, S\~{a}o Paulo, Brazil}
\author{Eduardo Miranda}
\affiliation{
Gleb Wataghin Physics Institute, The University of Campinas, 13083-859, Campinas, S\~{a}o Paulo, Brazil}

\date{\today}

\begin{abstract}
Finite-time driving of quantum systems generally produces excess energy above the quasistatic result. We study this problem for weakly driven one-dimensional critical systems in the Luttinger-liquid universality class. Combining linear response theory with bosonization, we obtain the asymptotic behavior of the excess work and show that the exponent controlling its decay is determined by the scaling dimension of the operator coupled to the source of work. We uncover a hierarchy of dynamical regimes in which this exponent increases continuously with the scaling dimension before saturating at universal plateau values. By engineering protocols with vanishing derivatives at the beginning and end of the control process, these plateaus can be systematically accessed, leading to 
asymptotic suppression of excess work. The resulting scaling applies to both global (bulk) and local control. We further support our analytical results with time-dependent density-matrix renormalization group calculations. Our findings establish a universal framework for understanding and optimizing finite-time control in quantum critical systems. 

\end{abstract}


\maketitle

Manipulating closed quantum systems in a finite time is generically an energy-consuming process \cite{Deffner_2020}. For  a non-equilibrium process of duration $\tau$,  the total average work $W(\tau)$ can be measured as the difference between the final and initial energies \cite{PRE_Minimal_work_2005,PRE_work_2007,PhyA_JARZYNSKI_2020}. The excess work consumed in this process is  defined as   
\begin{eqnarray}
\mathrm{W}_{\te{ex}}(\tau)\equiv W(\tau)-W_{\te{qs}}&\geq&0,
\end{eqnarray}
where $W_{\te{qs}}=W(\tau\to\infty)$ is the quasistatic work, i.e., the work associated with an infinitely slow process. Quantifying and minimizing $\mathrm{W}_{\te{ex}}(\tau)$ is a central challenge in quantum science, with direct implications for the energetic resources required to implement efficient new quantum technologies \cite{NJP_Brif_2010,RMP_AQC_2018,NP_Barzanjeh_2022,NP_King_2022}. Motivated by this problem,  we investigate the dependence of the excess work on the driving duration $\tau$ in a class of weakly driven one-dimensional (1D) critical systems  belonging to the Luttinger-liquid (LL) universality class \cite{giamarchi2004quantum}. Our results reveal how the asymptotic decay of $\mathrm{W}_{\te{ex}}(\tau)$ is affected by strong interactions, the protocol profile, and the nature of the control---global or local. We derive an asymptotic optimal control strategy for systematically suppressing $\mathrm{W}_{\te{ex}}(\tau)$ in such systems and show that the scaling dimension of the operator coupled to the source of work determines the maximum decay exponent of the excess work attainable through protocol engineering.

The question of how excess work scales with large but finite $\tau$  has been addressed in the literature in a variety of contexts, ranging from closed few- and many-body systems \cite{Polkovnikov_NP_2008,PRB_Pellegrini_2008,Eckstein_2010,PRL_Dora_2011,PRL_Bernier_2011,Soriani_2022,PRA_Soriani_2022,NJP_Rolandi_2023,PRL_Rolandi_2023,PRB_Dupays_2024,Cavalcante2024,Sela_PRL_2025} to open systems \cite{Scandi_Quantum_2019,Rolandi_2023}. Two theoretical frameworks have proven particularly successful in describing this scaling: the geometric approach \cite{PRL_Crooks_2012,Entropy_Matteo_2020,PRL_Van_2021} and adiabatic perturbation theory (APT) \cite{PRA_APT_2008}. The former predicts that $\mathrm{W}_{\te{ex}}(\tau)\sim\tau^{-1}$ for open systems in contact with a thermal bath \cite{PRL_Crooks_2012,Bonanca_JCP_2014}. In contrast, APT predicts $\mathrm{W}_{\te{ex}}(\tau)\sim\tau^{-2}$ for closed gapped systems  \cite{Soriani_2022}, where the finite energy gap is required for the validity of  the adiabatic theorem. 
For the class of gapless systems considered here,
low-energy excitations can invalidate a regular expansion in the
driving rate and produce nonanalytic excess-work scaling. Indeed, as we shall see below,  in certain regimes the exponent $\eta$ in $\mathrm{W}_{\te{ex}}(\tau)\sim\tau^{-\eta}$ can take any value between $1$ and $2$.

Luttinger liquids provide a paradigmatic example of interacting quantum systems in which conventional quasiparticle pictures and  mean-field approximations break down \cite{giamarchi2004quantum}. LL theory \cite{FDMHaldane_1981} describes the broad universality class of gapless 1D systems in which low-energy excitations have a linear dispersion relation. Within this theory, local perturbations excite collective bosonic modes, which govern the characteristic power-law decay of correlation functions. Well-established examples of LLs include electrons in quantum wires, spin chain materials, ultracold atomic gases, and edge states of topological phases \cite{Bouchoule2025}. Key aspects of  LL physics under out-of-equilibrium conditions have been investigated both theoretically \cite{PRL_Cazalilla_2006,PRA_Iucci_2009,PRL_Chamon_2011,PRA_Chamon_2013,PRB_Dora_2013,PRL_Schiro_2014,PRB_Schiro_2015,PRB_Dora_2019,Vir_PNAS_2020,PRL_Dora_2020,PRR_Haldar_2020,Sci_Datta_2023,Cavalcante2024,PRL_Flavia_2024,PRB_Dora_2024} and experimentally \cite{Hofferberth_NP_2007,RMP_LL_2011,Gring_Science_2012,Aimet2025,Yamazaki2025}. While the dynamics of LLs following sudden quenches has been extensively studied \cite{Mitra_2018}, their response to controlled finite-time driving remains relatively unexplored \cite{PRB_Dziarmaga_2011}, despite its relevance to quantum many-body state preparation in optical lattices \cite{PRL_Carcy_2021}. In this work, we address this question.

\paragraph*{Excess work for weakly driven systems.}
\label{Ew_LRT_sec}
Consider a closed quantum system at zero temperature described by a time-dependent Hamiltonian $\mc H(\lambda(t))=H_0+\lambda(t)\delta H$. Here, $H_0$ is an unperturbed Hamiltonian and  $\delta H$ describes the perturbation controlled  by an external agent which acts as the source of work. The time-dependent coupling $\lambda(t)$ is assumed to be a smooth function of $u\equiv t/\tau$ and varies from $\lambda(0)=0$ to $\lambda(1)=\lambda_{0}$. The precise form of the function $\lambda(u)$ specifies the protocol. Hereafter, we set $\hbar=1$.

We assume that for $t<0$ the system is prepared in the ground state of $H_0$, $|\Upsilon_0\rangle$. Provided that the coupling $\lambda(t)$ is weak compared to the effective energy scale of $H_0$, linear response theory (LRT) yields the excess work as a  functional of $\dot\lambda(u)=d\lambda(u)/du$  \cite{PRE_Bonanca_2018,PRL_2024_GLRP}
\begin{eqnarray}
    \mathrm{W}_{\te{ex}}(\tau)&=&\frac{1}{2}\int_0^1 dudu'\:\dot \lambda(u)\dot \lambda(u')\Psi(\tau(u'-u)),
    \label{excesswork}
\end{eqnarray}
with $u$ and $u'$ being dimensionless variables. The properties of the system are encoded in the relaxation function $\Psi(s)$  defined as  $\Psi(s)=-\int_0^{s}dt\:\Phi(t) +\te{const.}$, where $\Phi(t)= i\langle [\delta H(t),\delta H(0)]\rangle_0$ is the response function associated with the operator that appears in the perturbation \cite{Kubo_2}. Here, $\langle\cdots\rangle_0\equiv\langle\Upsilon_0|\cdots|\Upsilon_0\rangle$, and operators evolve in the interaction picture as $\delta H(t)=e^{iH_0t}\delta He^{-iH_0t}$. 
As the initial state is an eigenstate of $H_0$, the response function obeys $\Phi(t)=-\Phi(-t)$, which in turn implies $\Psi(s)=\Psi(-s)$. Finally, the constant term in the definition of $\Psi(s)$ is fixed to ensure that $\Psi(\infty)=0$ \cite{Kubo_2}.

\paragraph*{XXZ chain.}

As a concrete example, we take $H_0$ as the Hamiltonian for the XXZ spin chain
\begin{eqnarray}    H_0&=&J\sum_{j=1}^N\left(S^x_{j}S^x_{j+1} + S^y_{j}S^y_{j+1} + J_z S^z_{j}S^z_{j+1}\right).
    \label{XXZ_chain}
\end{eqnarray}
Here, $S^{a}_j$ with $a=x,y,z$ are spin-$1/2$ operators at site $j$,  we set $J\equiv 1$ as our energy unit, and $J_z$ is the anisotropy parameter. The nonequilibrium dynamics of the XXZ chain can be investigated in programmable quantum simulators \cite{Yao2021,Scholl2022,Dimitrova2023}. This model exhibits a gapless phase described by LL theory in the range $-1<J_z\leq 1$ \cite{giamarchi2004quantum}. In this phase, the decay of spin-spin correlations depends on the Luttinger parameter $K$, which is known exactly: $K=\frac{\pi/2}{\pi-\arccos(J_z)}$ \cite{giamarchi2004quantum,gogolin2004bosonization}. In particular, for $J_z=0$ the model can be mapped onto free fermions by a Jordan-Wigner transformation; at this point $K=1$. More generally, increasing $|J_z|$ corresponds to increasing the  interaction strength  in the fermionic representation.

For our purpose of controlling critical systems in a finite time, we assume that the driving does not push the system away from the LL universality class. This means that $\mc H(\lambda)$, for a fixed $\lambda\neq 0$, still has its equilibrium low-energy physics governed by LL theory. Clearly, this assumption does not hold for arbitrary operators $\delta H$. A renormalization group (RG) analysis of $\mc H(\lambda)$ reveals that this condition is satisfied whenever $\delta H$ is irrelevant in the RG sense \cite{sachdev1999quantum}. We therefore restrict our attention to  irrelevant operators $\delta H$. This condition also ensures the consistency of our perturbative approach, as it prevents the breakdown of Eq.~(\ref{excesswork}) due to higher-order contributions in $\lambda(t)$.

In the following, we distinguish between two   cases:   global (bulk) control and local control.

\paragraph*{Global control.}
\label{bulk_case_sec}
In this first case, we consider a perturbation that couples to the entire system, $\delta H=\sum_j O_j$, where $O_j$ is a local operator acting on site $j$. As an example, we choose $O_j=(-1)^j S^z_j$, which corresponds to  a staggered magnetic field along the $z$ direction. The equilibrium and nonequilibrium properties of the XXZ chain in the presence of a staggered field have been studied in the literature \cite{Kuz_PRB_2009,Huang_PRB_2013, Mendoza_PRE_2015, Vir_PNAS_2020,wildeboer2025}. In particular,  the staggered field is known to break the integrability of the XXZ chain \cite{Huang_PRB_2013}.

We use bosonization \cite{giamarchi2004quantum,Miranda_BJP_2003} to determine the regime in which the staggered field is an irrelevant perturbation. In the continuum limit, the unperturbed Hamiltonian   becomes $H_0\sim \frac{u_b}2\int_0^L dx\, [K(\partial_x\theta)^2+K^{-1}(\partial_x\phi)^2]$, where $u_b$ is the  velocity of the bosonic modes, $L$ is the chain length  with periodic boundary conditions, and $\theta(x)$ and $\phi(x')$ are bosonic fields that obey $[\partial_x\theta(x),\phi(x')]=-i{\delta(x-x')}$. Bosonization of the staggered-field term yields $\delta H\sim\frac{1}{\pi\alpha}\int_0^Ldx \cos[\sqrt{4\pi}\phi(x)]$, where $\alpha$ is a short-distance  cutoff of the order of the lattice spacing. This perturbation to the LL fixed point has scaling dimension $d_{\te{bulk}}=K$ and becomes irrelevant for $K>2$. Using the exact expression for the Luttinger parameter, we find that $\delta H$ is irrelevant in the regime $-1<J_z<-\frac1{\sqrt{2}}$.

Next, we calculate the relaxation function within LL theory. As explained in the  End Matter, we obtain
\begin{eqnarray}
    \Psi(t)&=&\frac{f(K)}{2\Lambda(K-1)}\te{Re}\left[\frac{1}{\left(1+i\Lambda t\right)^{2K-2}}\right],
    \label{rela_func_bulk}
\end{eqnarray}
where $\te{Re}[\cdots]$ denotes the real part, $\Lambda^{-1}=\alpha/u_b$  plays the role of a nonuniversal short-time scale, and $f(K)=\frac{L}{\alpha\pi^{3/2}}\frac{\Gamma(K-1/2)}{\Gamma(K)}$, and $\Gamma(x)$ is the gamma function. Note that $\Psi(t)$ decays as a power law with a $K$-dependent exponent, a hallmark of LL behavior \cite{giamarchi2004quantum}.

We can calculate the excess work $\mathrm{W}_{\te{ex}}(\tau)$ once  we specify the control protocol $\lambda(t)$.  We start with the linear protocol $\lambda(u)=\lambda_0u$. Within LRT, we require  $\lambda_0\ll J$. In the time interval $\Lambda^{-1}\ll\tau\ll L/u_b$, we obtain  
\begin{eqnarray}
    \mathrm{W}_{\te{ex}}(\tau)\simeq\frac{g(K)}{(\Lambda\tau)^2}+\mc O\left(\tau^{2-2K}\right),
    \label{exc_work_line_resul}
\end{eqnarray}
where $g(K)$ can be found in the End Matter. We see that the excess work decays as $\sim \tau^{-2}$, independently of $K$. Only the subleading term has an exponent that depends on $K$. This result is somewhat surprising, given that the decay of $\Psi(t)$ is controlled by $K$ [see Eq. (\ref{rela_func_bulk})].  Remarkably, the scaling $\mathrm{W}_{\rm ex}(\tau )\sim\tau^{-2}$ in Eq. (\ref{exc_work_line_resul}) coincides with  the APT predictions for gapped systems \cite{Soriani_2022}.
Thus a gapless system can display the same $\tau^{-2}$ power as a generic ramp in a gapped system.

\paragraph*{Generic protocol.}
\label{generic_prot_sec}
Let us now consider how a generic protocol affects the scaling of
$\mathrm{W}_{\te{ex}}(\tau)$. We use the following identity for the
time-dependent part of $\Psi(t)$,
\begin{equation}
    \te{Re}\left[\frac{1}{(1+i\Lambda t)^q}\right]
    =\frac{1}{\Gamma(q)}\int_0^\infty dx\:
    x^{q-1}e^{-x}\cos(\Lambda tx),
    \label{Gamma_relation}
\end{equation}
where $q=2K-2$. Substituting Eq.~(\ref{Gamma_relation}) into Eqs.~\eqref{rela_func_bulk} and (\ref{excesswork}), defining $y=\Lambda\tau x$ and $\mc J(y)=\int_0^1du\,\dot\lambda(u)e^{iyu}$, we obtain
\begin{equation}
    \mathrm{W}_{\te{ex}}(\tau)
    =\frac{h(q)}{(\Lambda\tau)^q}
    \int_0^\infty dy\:y^{q-1}e^{-y/(\Lambda\tau)}|\mc J(y)|^2,
    \label{excesswork2}
\end{equation}
where $h(q)=f(1+q/2)/(2\Lambda\Gamma(q+1))$. The boundary conditions $\lambda(0)=0$ and $\lambda(1)=\lambda_0$ imply $\mc J(0)=\lambda_0$.

Equation~(\ref{excesswork2}) shows that the large-$\tau$ behavior of $\mathrm{W}_{\te{ex}}(\tau)$ depends on whether the remaining integral converges
as $\Lambda\tau\to\infty$, which in turn is governed by the large-$y$ behavior of $\mc J(y)$. If $y^{q-1}|\mc J(y)|^2$ is integrable at infinity, we can set $e^{-y/(\Lambda\tau)}\simeq 1$ and directly obtain $\mathrm{W}_{\te{ex}}(\Lambda\tau\gg 1)\sim\tau^{-q}$. Otherwise, its dependence on $\tau$ can modify the decay exponent $\eta$.

Suppose first that at least one of the endpoint derivatives, $\dot\lambda(0)$ or $\dot\lambda(1)$, is nonzero. Integrating $\mc J(y)$ by parts once gives the large-$y$ result,
\begin{equation}
    \mc J(y)=\frac{\dot\lambda(1)e^{iy}-\dot\lambda(0)}{iy}
    +\mc O(1/y^2).
    \label{J}
\end{equation}
Since $q>2$, the large-$y$ tail of $y^{q-1}|\mc J(y)|^2$, which scales as $y^{q-3}$, is nonintegrable. Splitting the integral in
Eq.~(\ref{excesswork2}) at a fixed $y=y_0$, with $1\ll y_0\ll\Lambda\tau$, and using Eq.~(\ref{J}), we obtain, apart from the overall factor $h(q)$,
\begin{eqnarray}
    (\Lambda\tau)^q\mathrm{W}_{\te{ex}}
    &\sim& \int_{y_0}^\infty dy\:y^{q-3}e^{-y/(\Lambda\tau)}
    \left[A_1-B_1\cos(y)\right] \nonumber\\
    &&+\int_0^{y_0}dy\:y^{q-1}e^{-y/(\Lambda\tau)}|\mc J(y)|^2,
    \label{excesswork3}
\end{eqnarray}
where $A_1=[\dot\lambda(1)]^2+[\dot\lambda(0)]^2$ and
$B_1=2\dot\lambda(1)\dot\lambda(0)$. Since $\mc J(y)$ is bounded, the second integral is
$\mc O(1)$ for large~$\Lambda\tau$. In the first integral, the integrand is integrable at the origin. Changing variables back to $x=y/(\Lambda\tau)$, we can extend its lower limit to zero. It follows that the $A_1$ integral grows as $(\Lambda\tau)^{q-2}$, whereas the $B_1$ integral approaches a constant as $\Lambda\tau\to\infty$. The right-hand side of Eq.~(\ref{excesswork3}) therefore scales as $(\Lambda\tau)^{q-2}$. Thus, $\mathrm{W}_{\te{ex}}(\Lambda\tau\gg 1)\sim\tau^{-2}$ for any protocol with at least one nonzero endpoint derivative. This result generalizes the result in Eq. (\ref{exc_work_line_resul}).

\paragraph*{Engineered protocols.}
\label{eng_prot_sec}
Equation~(\ref{J}) connects the large-$y$ behavior of $\mc J(y)$ to the endpoint derivatives of $\lambda(u)$. This connection allows us to engineer protocols aimed at controlling the exponent $\eta$, as we now show.

Consider an engineered protocol satisfying
\begin{eqnarray}
    \lambda^{(n)}(0)=\lambda^{(n)}(1)=0,
    \qquad n=1,2,\ldots,k,
    \label{EP_condition}
\end{eqnarray}
where $\lambda^{(n)}(u)$ denotes the $n$-th derivative of $\lambda(u)$. We also require that at least one of $\lambda^{(k+1)}(0)$ and $\lambda^{(k+1)}(1)$ be nonzero. Integrating by parts $(k+1)$ times gives, for $y\gg 1$,
\begin{equation}
    \mc J(y)=
    \frac{(-1)^k\left(\lambda^{(k+1)}(1)e^{iy}
    -\lambda^{(k+1)}(0)\right)}{(iy)^{k+1}}
    +\mc O(1/y^{k+2}).
    \label{2}
\end{equation}
The large-$y$ tail of $y^{q-1}|\mc J(y)|^2$ now scales as
$y^{q-2k-3}$ and is integrable at infinity when $q<2k+2$. In the corresponding range of $K$, we conclude that
$\mathrm{W}_{\te{ex}}(\Lambda\tau\gg 1)\sim\tau^{2-2K}$.

\begin{figure}[t]
    \centering

    \begin{minipage}{0.48\columnwidth}
        \centering
        \includegraphics[width=\linewidth]{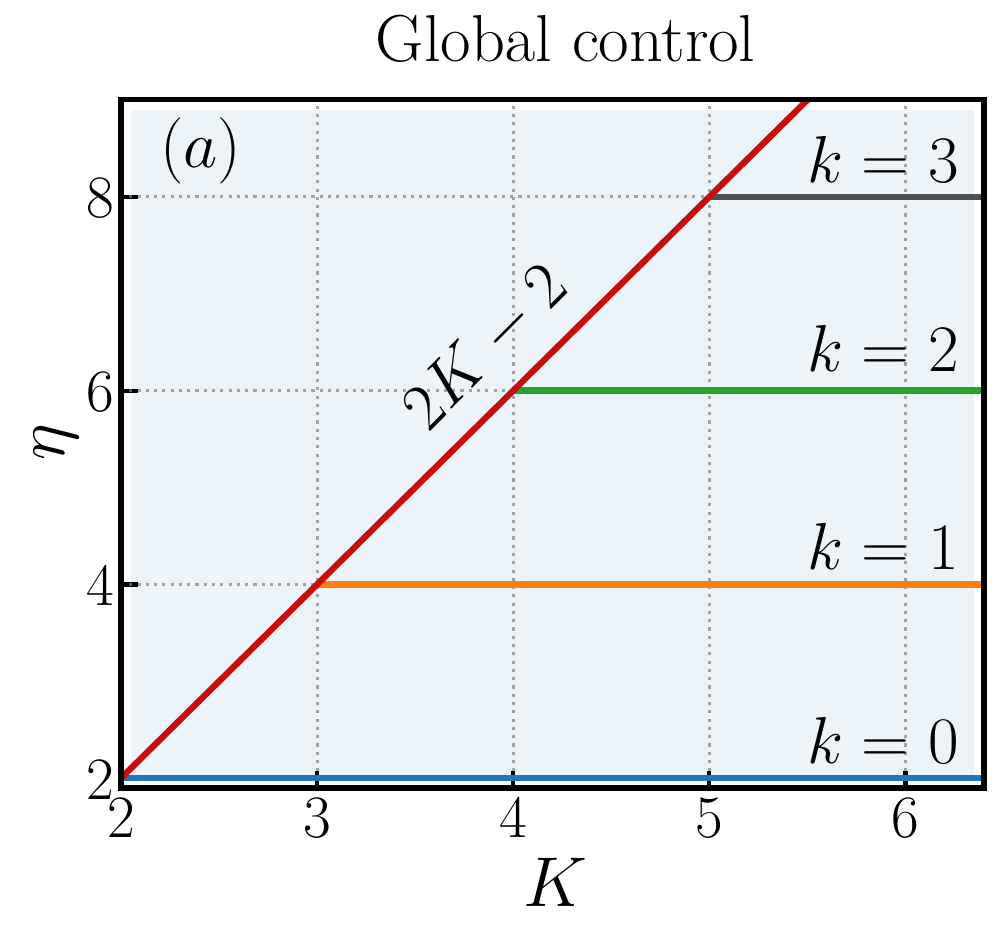}
    \end{minipage}
    \hfill
    \begin{minipage}{0.48\columnwidth}
        \centering
        \includegraphics[width=\linewidth]{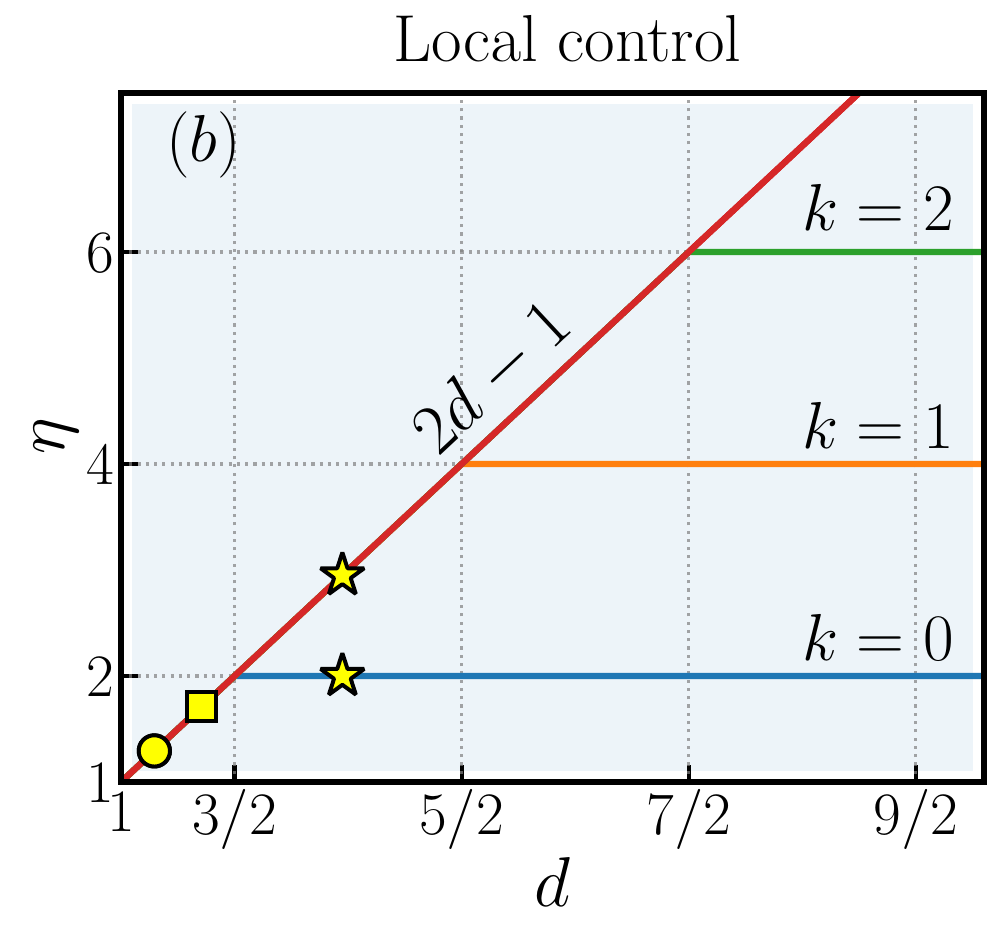}
    \end{minipage}

    \caption{Exponent $\eta$ governing the asymptotic behavior $\mathrm{W}_{\ma{ex}}(\Lambda\tau\gg1)\sim \tau^{-\eta}$ as a function of the scaling dimension of the perturbation. We consider two cases: (a)  global control, exemplified by a staggered magnetic field [see Eq. (\ref{Exc_work_In})];   (b) local control   [see Eq. (\ref{Exc_work_local})]. In both cases, the exponent varies with the scaling dimension (red lines) before reaching plateaus (shown in blue, orange, green, and black) that depend on the number  $k$ of vanishing derivatives at the endpoints of the protocol. The yellow symbols  in (b) represent the values of the exponents   extracted from tDMRG results, with symbols following the representation  in  Fig. \ref{fig_numerics}.}
    \label{fig_exponent}
\end{figure}

For $q\geq 2k+2$, we use the same split as in Eq.~(\ref{excesswork3}) (see the End Matter for details). For $q>2k+2$, the nonoscillatory tail contributes a term proportional to $(\Lambda\tau)^{q-2k-2}$ to the integral in Eq.~(\ref{excesswork2}), giving $\mathrm{W}_{\te{ex}}(\tau)\sim\tau^{-2k-2}$. The oscillatory integral gives an $\mc O(1)$ contribution. At $q=2k+2$, we have $\int_{y_0}^\infty dy\,e^{-y/(\Lambda\tau)}/y
=\ln(\Lambda\tau)+\mc O(1)$, while the cosine integral approaches a constant. Hence $\mathrm{W}_{\te{ex}}(\tau)\sim
\tau^{-2k-2}\ln(\Lambda\tau)$. 

It is worth stressing that the authors of Ref. \cite{Eckstein_2010} have already demonstrated that the smoothness of the protocol can modify the asymptotic behavior of the excess work.

Combining the previous results, we obtain that the exponent $\eta$ for such engineered protocols has the intricate behavior
\begin{eqnarray}
    \eta&=&\te{min}(2K-2,2k+2).
    \label{Exc_work_In}
\end{eqnarray}
The transition from a $K$-dependent exponent to a constant exponent
occurs at $K_k^*=k+2$, where a logarithmic correction
$\ln(\Lambda\tau)$ appears. In terms of the anisotropy parameter, the
transition occurs at $J_{z,k}^*=-\cos\left(\frac{\pi}{2k+4}\right)$.
For a generic protocol, $k=0$, and $\eta=2$ throughout the regime
considered. Equation~(\ref{Exc_work_In}) is the main result of this work.

In Fig. \ref{fig_exponent}(a), we schematically illustrate the behavior of $\eta$ according to Eq. (\ref{Exc_work_In}). The exponent increases continuously as $\eta=2K-2$ until it reaches a certain plateau, where it remains pinned to a value that depends only on the number $k$ of vanishing derivatives in Eq. (\ref{EP_condition}). We stress that, for all engineered protocols with $k\geq1$, the exponent  is always greater than $2$. Therefore, compared with a generic protocol [see  Eq. (\ref{exc_work_line_resul})], the engineered protocol suppresses the excess work more efficiently at large $\tau$, thereby reducing the energetic cost for the external agent.

The behavior shown in Fig. \ref{fig_exponent}(a) can be generalized to arbitrary global controls that involve irrelevant operators in critical systems. Within LL theory, the essential input is the scaling dimension of the perturbation, which governs the power-law decay of the response function. In the example of a staggered magnetic field, the exponent $\eta=2K-2$ for $2<K<K^*_k$ stems from the correlation function for the operator $\cos[\sqrt{4\pi}\phi(x)]$, which has scaling dimension $K$. In general, if the coupling between the system and the external source of work is described by a generic operator $\delta H=\sum_jO_j$ where $O_j$ has scaling dimension $d_{\te{bulk}}$, the excess work is expected to scale as $\tau^{2-2d_{\te{bulk}}}$ for $2<d_{\te{bulk}}<k+2$.

\paragraph*{Local control.}
\label{local_case_sec}
For this second case, we consider local perturbations of the form $\delta H=O_j$, where $O_j$ acts on a finite number of sites near site $j$. The restriction to local correlations makes this case quantitatively different from the previous one. In the continuum limit, we represent $O_j$ by an operator $O(x)$ written in terms of the bosonic fields in the LL theory. For local perturbations, the operator $O(x)$ is irrelevant if its scaling dimension $d$ is greater than $1$. Conformal invariance determines the autocorrelation function to be of the form $\langle O(x,t)O(x,0)\rangle_0\propto(1+i\Lambda t)^{-2d}$ \cite{gogolin2004bosonization}, from which we obtain   
\begin{eqnarray}
    \Psi(t)&=&\frac{2}{\Lambda(2d-1)}\te{Re}\left[\frac{1}{\left(1+i\Lambda t\right)^{2d-1}}\right].
    \label{rela_func_local}
\end{eqnarray}
We find that $\Psi(t)$ now decays as $t^{1-2d}$, in contrast to the  $t^{2-2d_{\te{bulk}}}$ decay observed for the bulk perturbation [see Eq. (\ref{rela_func_bulk})]. The difference in the exponents arises  from the additional spatial integration in the bulk case. In effect, the exponent  $\eta$ for the local control case can be obtained from the bulk result by making the substitution $d_{\te{bulk}}\to d+1/2$.

Similarly to the result in Eq. (\ref{Exc_work_In}), for the local control case we find that the excess work decays as a power law with exponent  
\begin{eqnarray}
    \eta&=&\te{min}(2d-1,2k+2).
    \label{Exc_work_local}
\end{eqnarray}
The transition between $d$-dependent and constant exponents is located at $d^*_k=k+3/2$, where a $\ln{(\Lambda \tau)}$ correction appears. 
Unlike the bulk case, local control admits an interaction-dependent regime even for $k=0$, before reaching the $\eta=2$ plateau, see Fig. \ref{fig_exponent}(b). Thus, even for a non-engineered protocol, there exists a region ($1<d<3/2$) in which the exponent varies  continuously as $\eta=2d-1$. Remarkably, in this region  $\eta$ takes its lowest value among  all cases considered, with $1<\eta<2$.

Equations (\ref{Exc_work_In}) and (\ref{Exc_work_local}) reveal that, for both global and local control, the scaling dimension of the perturbing operator dictates the fastest asymptotic decay of the excess work attainable through protocol engineering. Beyond this limit, increasing the protocol smoothness yields no further improvement in the decay exponent.

\begin{figure}[t]
    \centering
\includegraphics[width=1.0\linewidth]{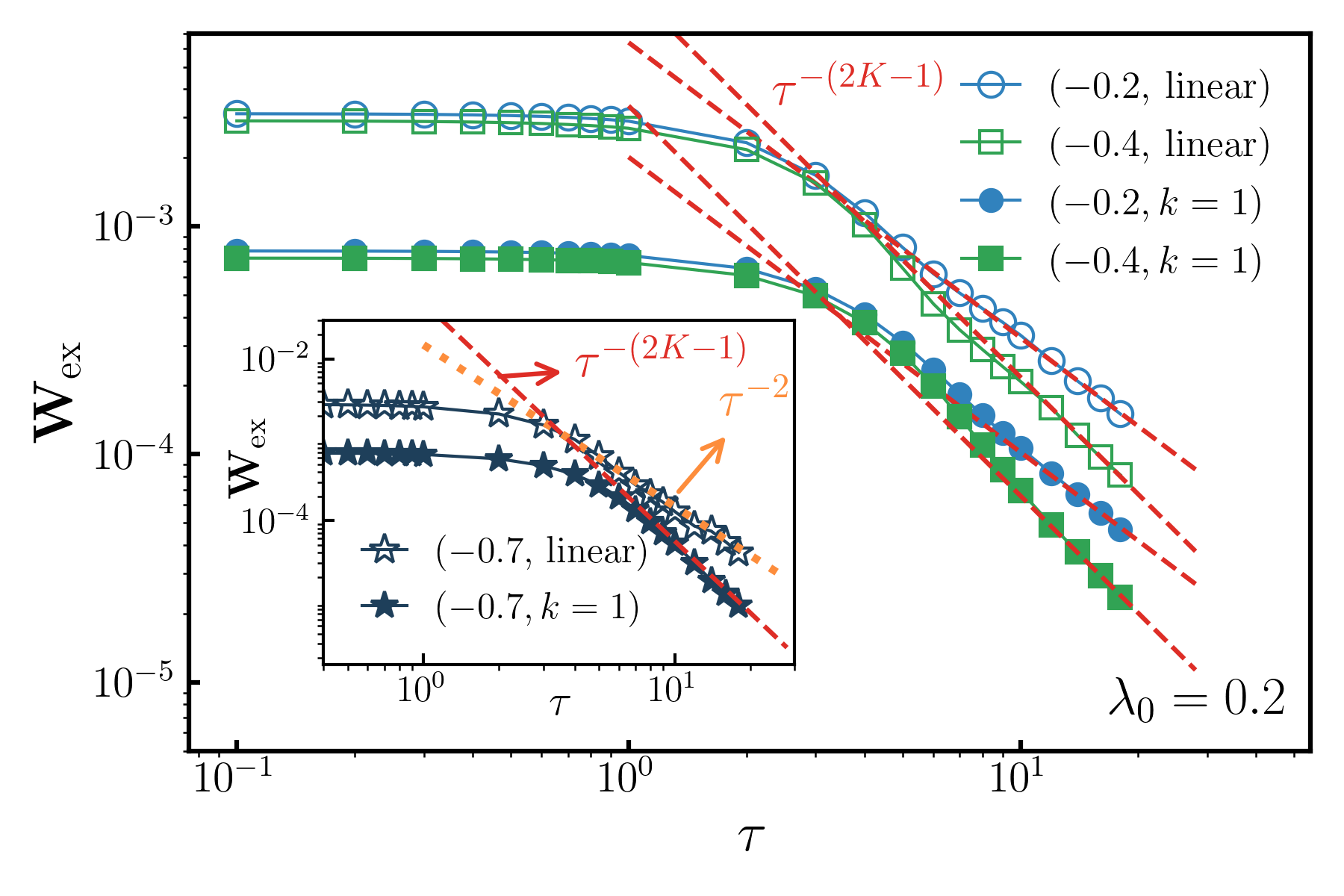}
    \caption{tDMRG results for the excess work $\mathrm{W}_{\mathrm{ex}}(\tau)$ associated with a local control $\delta H=S^x_jS^x_{j+1}+S^y_jS^y_{j+1}$, as a function of $\tau$ for different values of $J_z$ and protocols, identified in the legend by $(J_z,\text{protocol})$. Here we fix $J=1$ and $\lambda_0=0.2$. For clarity, the data represented by open symbols have been rescaled by a factor of $1/4$. Dashed lines indicate power-law decays with exponent $\eta=2K-1$. The inset shows the results for $J_z=-0.7$, where the dotted line represents a power-law decay with exponent $\eta=2$. Note that, for $k=1$ and $J_z=-0.7$, the leading decay exponent is interaction dependent and greater than 2.} 
    \label{fig_numerics}
\end{figure}

\paragraph*{Numerical results.}
\label{sec_Nume_res}
To test our analytical predictions, we have performed numerical simulations using the time-dependent density-matrix renormalization group (tDMRG) method \cite{White2004}. Since both global and local control protocols exhibit the predicted plateau behavior, we focus on local control, which allows us to access longer protocol durations $\tau$. Specifically, we simulate the XXZ chain in Eq. (\ref{XXZ_chain}) with a time-dependent interaction on a single  bond, described by $\delta H=S^x_jS^x_{j+1}+S^y_jS^y_{j+1}$. For $-\frac1{\sqrt{2}}<J_z<0$, corresponding to  $1<K<2$, the leading contribution in its bosonized form is $\delta H\sim \sin[\sqrt{4\pi}\phi(j)]$, which has scaling dimension $d=K$ (see the End Matter).  In our simulations, we used a chain of $N=100$ sites with open boundary conditions, took $j=50$, and retained up to 300 states per DMRG block, which yields ground-state energies with an accuracy better than $10^{-7}$. The time evolution is computed using a second-order Suzuki-Trotter decomposition with time step $\delta t=0.01$.

In Fig. \ref{fig_numerics}, we show $\mathrm{W}_{\te{ex}}(\tau)$ for three values of $J_z$. For the linear protocol, $\lambda(u)=\lambda_0u$, the results follow the predicted  exponent $\eta=2K-1$ for $J_z=-0.2$ and $-0.4$, corresponding to  $K=1.147$ and $1.355$, respectively. For $J_z=-0.7$, corresponding to $K=1.975>3/2$, the exponent lies in the first plateau, characterized by $\eta=2$ (see the inset). In contrast, for the $k=1$ protocol, $\lambda(u)=\lambda_0(3u^2-2u^3)$, the exponent continues to follow $\eta=2K-1$ even for $J_z<-0.5$ ($K>3/2$), see the inset. Therefore, the results for $J_z=-0.7$ confirm the presence of the first plateau. Additional numerical results for other values of $J_z$ and $\lambda_0$, presented in the End Matter, further corroborate our analytical predictions.

\paragraph*{Concluding remarks.}
\label{sec_conclu}
In summary, we have established how the excess work in weakly driven LLs depends on interactions and the control protocol. For both global and local couplings to an external agent, we found that the decay exponent of $\mathrm{W}_{\ma{ex}}(\tau)$ exhibits a transition between a regime in which it increases linearly with the scaling dimension of the perturbation and a regime characterized by interaction-independent plateaus. These analytical findings are consistent with numerical calculations. By engineering  protocols with vanishing derivatives at the beginning and at the end of the control, one can access higher plateaus, thereby  accelerating the decay of the excess work and reducing the energetic cost of the process. Interestingly, this optimization strategy had so far been justified only for gapped systems where adiabatic perturbation theory leads to a $1/\tau$-expansion of the excess work that explicitly depends on the derivatives at the end points of the protocol \cite{Soriani_2022}. Our linear-response approach extends this strategy to gapless systems. Finally, our findings establish a direct connection between the scaling dimension of operators in quantum critical systems and the ultimate asymptotic suppression of excess work achievable through protocol engineering.

\paragraph{Acknowledgements}
We acknowledge the financial support from the S\~{a}o Paulo Research Foundation (FAPESP), Brazil, Process No. 2022/15453-0, from Finep (Grant No. 1699/24 IIFFINEP, R.G.P.), and National Council for Scientific and Technological Development (CNPq) (Grants Nos.
309569/2022-2 and 404274/2023-4, R.G.P., Grant No. 304120/2022-7, M.V.S.B., Grant No. 309720/2025-5, E.M., and Grant No. 446378/2024-0, F.B.R.). We acknowledge the use of ChatGPT (GPT-6.2, OpenAI) as an auxiliary tool for verifying the analytical results presented in this work.

\bibliography{references}

\section*{End Matter}

\section{Bosonization details}

Spin-$1/2$ operators can be fermionized using the Jordan-Wigner transformation \cite{sachdev1999quantum}, $S^{+}_j=(-1)^je^{i\pi\sum_{n<j}c^{\dagger}_nc^{\phantom\dagger}_n}c^{\dagger}_j$ and $S_j^z=c_j^\dagger c_j-1/2$, where $c_j$ are fermionic annihilation operators. In the continuum theory describing the low-energy sector of $H_0$, the lattice fermion operator can be written as $c_j\sim e^{ik_Fx}\psi_R(x)+e^{-ik_Fx}\psi_L(x)$, where $\psi_{R/L}(x)$ are  right- and left-moving chiral fermions \cite{giamarchi2004quantum} and $k_F$ is the Fermi wave vector. For the XXZ chain with zero magnetization,  $k_F=\pi/2$  in units where the lattice spacing is set to unity. 

The bulk perturbation associated with  a staggered field, $\delta H=\sum_j(-1)^jS_j^z$, involves the operator $S_j^z$. In the continuum limit, we have 
\begin{equation}
S^z_j\sim\rho_R(x)+\rho_L(x)+(-1)^j[\psi^{\dagger}_R(x)\psi^{\phantom\dagger}_L(x) + \te{H.c.}],
    \label{Sz}
\end{equation}
where $\rho_{R/L}=:\psi^{\dagger}_{R/L}\psi_{R/L}:$, with  $::$ denoting normal ordering. As a consequence, the bulk perturbation  becomes 
\begin{equation}
    \delta H\sim\int_0^Ldx\:\psi^{\dagger}_R(x)\psi_L(x) + \te{H.c.},
    \label{Staggered1}
\end{equation} 
where we drop the rapidly oscillating terms in the integral. We bosonize the chiral fermions as $\psi_{R/L}(x)\sim (2\pi\alpha)^{-1/2}e^{-i\sqrt{2\pi}\phi_{R/L}(x)}$ \cite{giamarchi2004quantum,Miranda_BJP_2003}. Here, $\phi_{R/L}(x)$ are chiral bosonic fields related to $\theta(x)$ and $\phi(x)$ introduced in the main text by $\phi_{L,R}(x)=[\theta(x)\pm\phi(x)]/\sqrt2$. Using this bosonization prescription, we obtain 
\begin{equation}
    \delta H\sim\frac{1}{\pi\alpha}\int_0^Ldx\:\cos[\sqrt{4\pi}\phi(x)].
    \label{staggered2}
\end{equation}

For a local control, we consider the perturbation by the bond operator $O_j=S^x_jS^x_{j+1}+S^y_jS^y_{j+1}$ for a given site $j$. In terms of fermions in the continuum limit, we have
\begin{eqnarray}
  O_j&\sim&-\frac{i}2\left[\psi^{\dagger}_R(x)\partial_x\psi^{\phantom\dagger}_R(x)-\psi^{\dagger}_L(x)\partial_x\psi^{\phantom\dagger}_L(x)-\text{H.c.}\right]\nonumber\\
    &&+i(-1)^{j}\left[\psi^{\dagger}_R(x)\psi_L(x)-\te{H.c.}\right].
    \label{bond_op}
\end{eqnarray}
Upon bosonization, we obtain \bea
O_j&\sim& \mc A[(\partial_x\phi)^2+ (\partial_x\theta)^2]\nonumber\\
&&+(-1)^j\mc B\sin[\sqrt{4\pi}\phi(x)],
\eea
where $\mc A$ and  $\mc B$  are prefactors. The uniform part of $O_j$ involves the  quadratic operators that appear in the energy density of the LL Hamiltonian and has scaling dimension $d=2$.  The staggered part of $O_j$ can be identified with the dimerization operator and has scaling dimension $d=K$. In the regime where $1<K<2$, the dimerization operator has the lowest scaling dimension. To obtain the dominant contribution to the excess work in this case, we retain only  $O_j\sim \sin[\sqrt{4\pi}\phi(x)]$.

\begin{figure}
    \centering
    \includegraphics[width=1\linewidth]{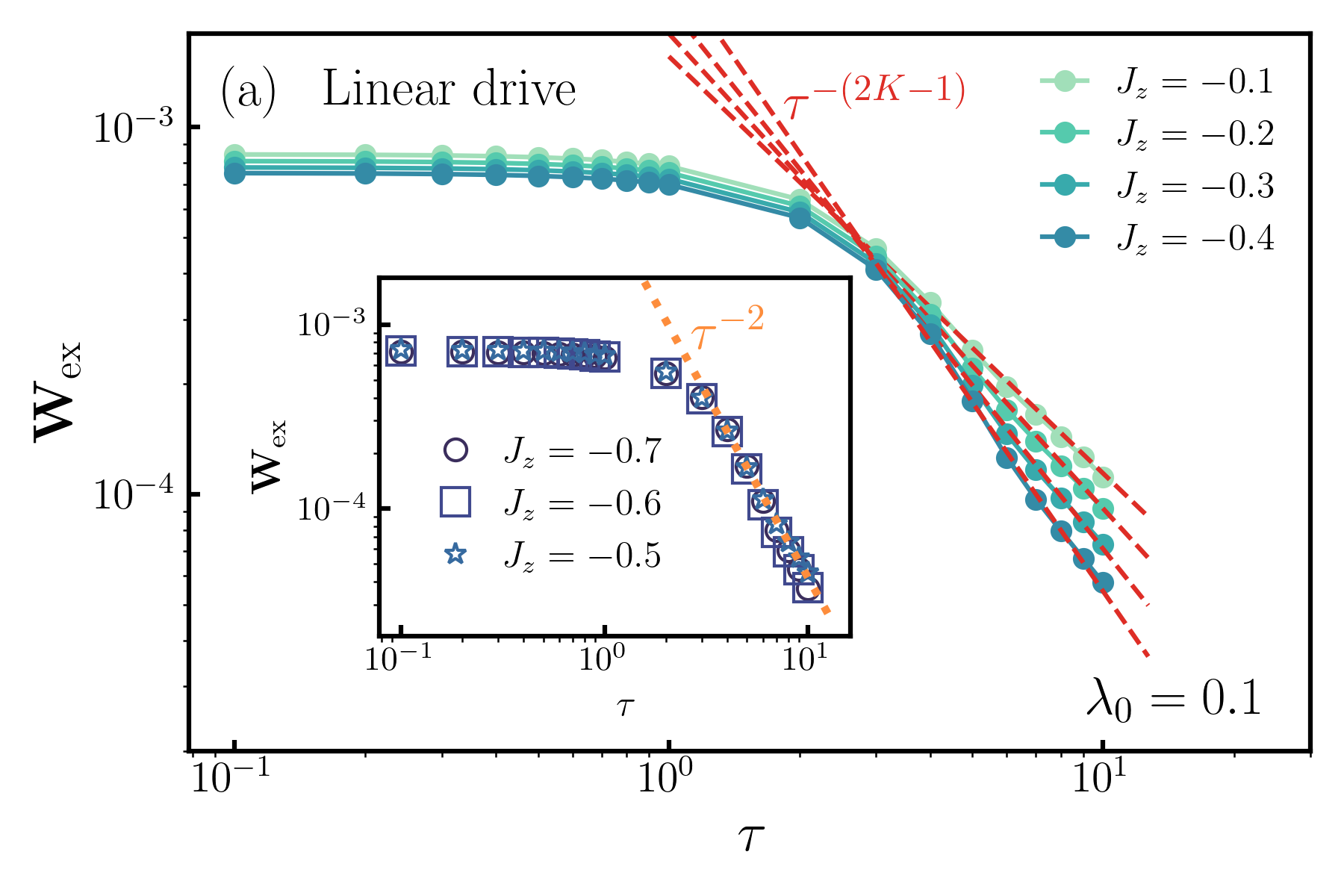}
        \includegraphics[width=1\linewidth]{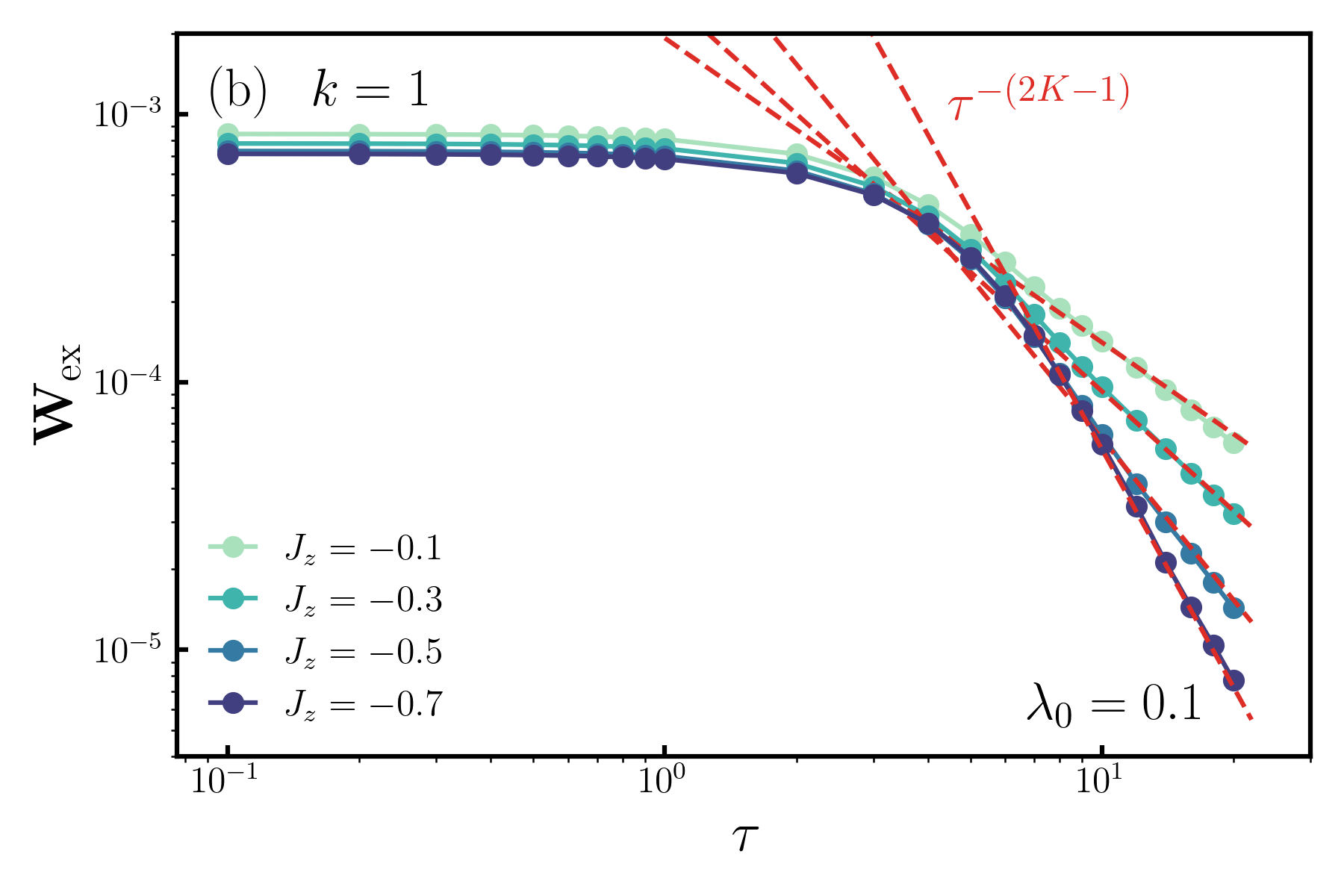}

    \caption{Excess work $\mathrm{W}_\text{ex}(\tau)$ as a function of $\tau$ for $\lambda_0=0.1$, using (a) the linear and (b) the $k=1$ protocol. The dashed lines indicate $\tau^{-(2K-1)}$, and the dotted line in the inset of (a) indicates $\tau^{-2}$.}
    \label{fig:dmrg_01}
\end{figure}

\section{Relaxation function}

To calculate the response function $\Phi(t)$ in the bulk case, we need the ground-state correlator of the operator $O_{\rm bulk}(x)\equiv\cos[\sqrt{4\pi}\phi(x)]$. Using the bosonization rules, we obtain 
\begin{eqnarray}
  &&  \langle O_{\rm bulk}(x,t)O_{\rm bulk}(x',t') \rangle_0\nonumber\\
&=&    \frac{1/2}{\left[1+\left(\Delta x/\alpha\right)^2-\left(\Lambda\Delta t\right)^2+2i\Lambda\Delta t\right]^{K}},
\end{eqnarray}
where $\Delta x =x-x'$ and $\Delta t=t-t'$. From this result, it is immediately seen that the scaling dimension of $O_{\rm bulk}(x)$ is $d_{\rm bulk}=K$. Calculating the commutator and subsequently performing the spatial integration, we arrive at
\be
    \Phi(t)=\frac{L}{\alpha\pi^{3/2}}\frac{\Gamma(K-\frac12)}{\Gamma(K)}\frac{\sin\left[(2K-1)\arctan\Lambda t\right]}{\left[1+(\Lambda t)^2\right]^{\frac{2K-1}{2}}}.
    \label{resp_func_bulk}
\ee
Integrating Eq. (\ref{resp_func_bulk}), we obtain the relaxation function $\Psi(t)$ in Eq. (\ref{rela_func_bulk}). Performing the integrals   in Eq. (\ref{excesswork}) for the linear protocol with $\lambda(u)=\lambda_0u$, we obtain the result for the excess work in Eq. (\ref{exc_work_line_resul}), with the prefactor\bea
g(K)&=&\frac{L\lambda_0^2}{\alpha\pi^{3/2}}\frac{\Gamma(K-1/2)}{\Gamma(K)}\frac{1}{2\Lambda(K-1)}\nonumber\\
&&\times\frac{1}{2(K-2)(2K-3)}.
\eea

In the local-control case, we calculate the response function $\Phi(t)$ for the operator  $O(x)\sim\sin[\sqrt{4\pi}\phi(x)]$. Up to a multiplicative factor, we obtain  
\begin{eqnarray}
\Phi(t)=\frac{2\sin\left[2d\arctan\left(\Lambda t\right)\right]}{\left[1+(\Lambda t)^2\right]^{d}}.
\label{resp_func_local}
\end{eqnarray}
Integrating this result, we obtain the relaxation function in Eq. (\ref{rela_func_local}).

\section{Large-$\tau$ behavior of $\mathrm{W}_{\te{ex}}(\tau)$ for engineered protocols}

For $q\geq 2k+2$, the integral in Eq.~(\ref{excesswork2}) diverges as $\Lambda\tau\to\infty$. Splitting the integral at a fixed $y=y_0\gg 1$, as in Eq.~(\ref{excesswork3}), and omitting the overall factor $h(q)$, we obtain 
\begin{equation}
\begin{aligned}
    &(\Lambda\tau)^q\mathrm{W}_{\te{ex}}(\Lambda\tau\gg 1)\\
    &\quad\sim \mc O(1)
    +A_k\int_{y_0}^\infty dy\:
    y^{q-2k-3}e^{-y/(\Lambda\tau)}\\
    &\qquad
    -B_k\int_{y_0}^\infty dy\:
    y^{q-2k-3}e^{-y/(\Lambda\tau)}\cos(y).
\end{aligned}
\label{excesswork4}
\end{equation}
The $\mc O(1)$ term comes from the interval $0\leq y\leq y_0$, in which $\mc J(y)$ is bounded. The coefficients are
$A_k=[\lambda^{(k+1)}(1)]^2+[\lambda^{(k+1)}(0)]^2$ and
$B_k=2\lambda^{(k+1)}(1)\lambda^{(k+1)}(0)$.

Setting $x=y/(\Lambda\tau)$, we can rewrite the last two terms in Eq.~(\ref{excesswork4}) as
\begin{equation}
\begin{aligned}
    &A_k(\Lambda\tau)^{q-2k-2}
    \int_{y_0/(\Lambda\tau)}^\infty dx\:
    x^{q-2k-3}e^{-x}\\
    &-B_k(\Lambda\tau)^{q-2k-2}
    \int_{y_0/(\Lambda\tau)}^\infty dx\:
    x^{q-2k-3}e^{-x}\cos(\Lambda\tau x).
\end{aligned}
\label{excesswork5}
\end{equation}
For $q> 2k+2$ and $y_0\ll\Lambda\tau$, we can send the lower limit to zero in both integrals. It follows that the first integral is $\Gamma(q-2k-2)$. The second integral, due to the oscillations introduced by the cosine, decays as $(\Lambda\tau)^{2k+2-q}$. Thus, we conclude that the right-hand side of Eq. (\ref{excesswork4}) grows as $(\Lambda\tau)^{q-2k-2}$, giving $\mathrm{W}_{\te{ex}}(\tau)\sim\tau^{-2k-2}$.

For $q=2k+2$, the integrand of the first integral in Eq. (\ref{excesswork5}) diverges logarithmically at the origin: $\int_{y_0/(\Lambda\tau)}^\infty dx\,e^{-x}/x
=\ln(\Lambda\tau)+\mc O(1)$. The oscillatory integrand of the second integral leads to $\int_{y_0/(\Lambda\tau)}^\infty dx\,e^{-x}\cos(\Lambda\tau x)/x =\mc O(1)$. Thus, $\mathrm{W}_{\te{ex}}(\tau)\sim \tau^{-2k-2}\ln(\Lambda\tau)$.

\section{Further Numerical Results}

Here, we present additional  tDMRG results for the excess work for the local control in the XXZ chain. Figure \ref{fig:dmrg_01} shows the excess work for the (a) linear protocol and (b) $k=1$  protocol given by $\lambda(u)=\lambda_0(3u^2-2u^3)$. Here we fix $\lambda_0=0.1$. For the linear protocol, the exponent varies with the interaction strength for $-0.5<J_z<0$ and remains stuck at the $\eta=2$ plateau for $J_z<-0.5$; see the inset of Fig. \ref{fig:dmrg_01}(a). For the engineered protocol with $k=1$, the exponent $\eta$ is determined by the interaction strength throughout the investigated range $J_z\in[-0.7,-0.1]$. The exponent is consistent with the analytical prediction $\eta=2K-1$.

\end{document}